\documentclass[journal]{IEEEtran}
\usepackage{url}
\usepackage[center]{caption}

\usepackage{float}
\usepackage{placeins}

\ifCLASSINFOpdf
\else
\fi
\usepackage{graphicx}

\begin{document}
%
\title{An RTL Implementation of the Data Encryption Standard (DES)}
%
%
%

\author{Ruby Kumari,~\IEEEmembership{Student Member,~IEEE,}
       Jai Gopal Pandey,~\IEEEmembership{Senior Member,IEEE,}
        and~Abhijit Karmakar,~\IEEEmembership{}
\thanks{Manuscript received December 19, 2022}}

\maketitle

\begin{abstract}
Data Encryption Standard (DES) is based on the Feistel block cipher, developed in
1971 by IBM cryptography researcher Horst Feistel. DES uses 16 rounds of the Feistel
structure. But with the changes in recent years, the internet is starting to be used more
to connect devices to each other. These devices can range from powerful computing
devices, such as desktop computers and tablets, to resource constrained devices, When
it comes to these constrained devices, using a different key for each round cryptography
algorithms fail to provide necessary security and performance.

\end{abstract}


%
\IEEEpeerreviewmaketitle

\section{Introduction}
%
%
%
%
\IEEEPARstart{T}{his} Security is a prevalent concern in information and data systems of all types. Historically, military and national security issues drove the need for secure communications.
Recently, security issues have pervaded the business and private sectors. E-commerce
has driven the need for secure internet communications. Many businesses have fire-
walls to protect internal corporate information from competitors. In the private sector,
personal privacy is a growing concern. Products are available to scramble both e-mail
and telephone communications. One means of providing security in communications is
through encryption. By encryption, data is transformed in a way that it is rendered
unrecognizable. Only by decryption can this data be recovered. Ostensibly, the process
of decryption can only be performed correctly by the intended recipients. The validity of
this assertion determines the “strength” or “security” of the encryption scheme. Many
communications products incorporate encryption as a feature to provide security. This
application report studies the implementation of one of the most historically famous and
widely implemented encryption algorithms, the Data Encryption Standard (DES). The
Data Encryption Standard is a symmetric-key block cipher published by the National
Institute of Standards and Technology (NIST) for the encryption of digital data. DES
is probably one of the best-known cryptographic algorithms and has been widely used
since its introduction in 1976. Although its short key length of 56 bits makes it too inse-
1
cure for applications, it has been highly influential in the advancement of cryptography.
The DES must be stronger than the other cryptosystems in security. The goal of
this project is to develop a python code for SDES and DES. Before building our design,
we need an overview of cryptography, followed by a description of the DES algorithm.


 



\subsubsection{Overview of Cryptography}
 Cryptography is a type of rule or technique by which
private or sensitive information is secured from the public or other members. It plays
a vital role in preserving data integrity, confidentiality and user privacy. An encryption
algorithm can convert imported essential data to encrypted data (plaintext into cipher-
text). This data would be of no use to a person that does not possess the encryption
key. The use of Cryptography in passwords is a very famous example. Cryptography is
based on mathematical theory and some Computer Science principles. There are many
terminologies related to cryptography. Some terms are defined below.

• Ciphertext: Conversion of plain text into intelligible text is called ciphertext.

• Cipher: It is a technique of encryption and decryption. Critical and algorithms
play vital role in this technique.

• Symmetric: It is a kind of cryptosystem. It uses same key for encryption and
decryption. It is faster than asymmetric.

• Asymmetric: It is also a kind of cryptosystem. It uses a public key for the
encryption and a private key for the decryption of any message.

• Cryptanalysis: It studies cracking the encryption of the algorithms.

\subsection{Symmetric Ciphers Model}
Symmetric-key (or private-key) encryption can be simply illustrated with the schematic
shown in Figure 1.

\par


\begin{figure}[ht]
     \centering
 \includegraphics[width=0.5\textwidth]{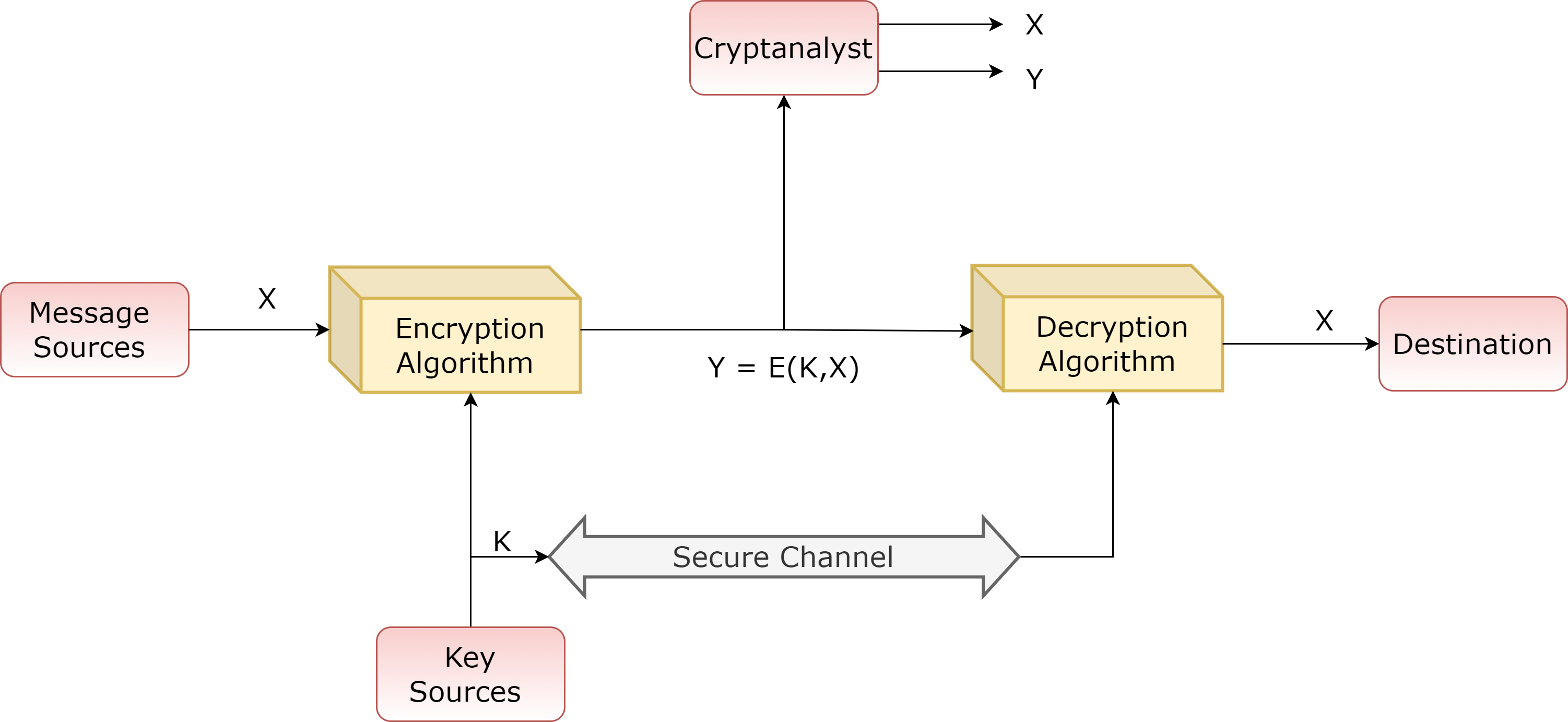}
\caption{Symmetric Cryptosystem model.}
\label{FIG:Symmetric Cryptosystem model} 
\end{figure}

A symmetric encryption scheme has five main parts, that is,

\begin{itemize}
\item \textbf{Encryption algorithm}: The encryption algorithm performs various substitutions and transformations on plaintext.
\item \textbf{Secret key}: The secret key is also input to the encryption algorithm. The key is a value independent of the plaintext and the algorithm. The algorithm will produce a different output depending on the specific key. The exact substitutions and transformations performed by the algorithm depend on the key.
\item \textbf{Ciphertext}: This is the scrambled message produced as output. It depends on the plaintext and the secret key. For a given message, two different keys will produce two different ciphertexts. The ciphertext is an random stream of data.
\item \textbf{Decryption algorithm}: This is essentially the encryption algorithm run in reverse. It takes the ciphertext and the secret key and produces the original plaintext.
\end{itemize}

Alice and Bob want to communicate over an un-secure channel, but Oscar is trying to read the message. So Alice and Bob must use a cryptosystem to prevent Oscar from reading the message.
Let us take a closer look at the essential elements of a symmetric encryption scheme using Figure 1. A source produces a message in plaintext, $X = [X1, X2, \ldots, XM]$. The \textit{M} elements of \textit{X} are letters in some finite alphabet. Traditionally, the alphabet usually consisted of t6 capital letters. Nowadays, the binary alphabet {0, 1} is typically used. For encryption, a key of the form $K = [K1, K2, ……., KJ]$ is generated. If the key is generated at the message source, then it must also be provided to the destination using some secure channel. Alternatively, a third party could generate key and securely deliver it to both source and destination.
The encryption algorithm forms the ciphertext as given in \ref{eq}.
\begin{equation}
        Y = [Y1,Y2,....,YN]
        \label{eq}
\end{equation}

with the message \textit{X} and the encryption key \textit{K} as it. We can write this as given in \ref{eqq}. 
\begin{equation}
    Y = E(K,X)
    \label{eqq}
\end{equation}

This notation indicates that \textit{Y} is produced by using encryption algorithm \textit{E} as a function of the plaintext \textit{X}, with the specific process determined by the value of the key \textit{K}.
The intended receiver, in possession of the key, can invert the transformation:
$X = D(K, Y)$.
An opponent, observing Y but not having access to \textit{K} or \textit{X}, may attempt to recover \textit{X} or \textit{K} or both \textit{X} and \textit{K}. It is assumed that the opponent knows the encryption (E) and decryption (D) algorithms. If the opponent is interested in only this particular message, then the focus of the effort is to recover \textit{X} by generating a plaintext estimate \textit{X}. Often, however, the opponent is interested in being able to read future messages as well, in which case an attempt is made to recover \textit{K} by generating an estimate \textit{K}.

\subsection {Simplified Data Encryption Standard}

The S-DES encryption algorithm takes an 8-bit block of plaintext and a 10-bit key as input and produces an 8-bit block of ciphertext as output. The S-DES decryption algorithm takes an 8- bit block of ciphertext and the same 10-bit key used to produce that ciphertext as input and produces the original 8-bit block of plaintext.
Simplified DES (SDES) was designed for educational purposes only, to help students learn about modern cryptanalytic techniques \cite{lu2022encryption}. SDES has similar properties and structure as DES but has been simplified to make it much easier to perform encryption and decryption by hand with
 
Pencil and paper. Some people feel that learning SDES gives insight into DES and other block ciphers, and insight into various cryptanalytic attacks against them.

An adversary trying to interrupt two communicating parties may have one of the four main goals:
\begin{enumerate}
\item Read the secret message.
\item Find the secret key, so that they can read all messages encrypted with that key.
\item Modify the message sent by Alice and go unnoticed by both parties.
\item Act like Alice and send a message to Bob, to make Bob think he is communicating with Alice when in reality he is communicating with the adversary.
\end{enumerate}
%

\begin{figure}[]
\centering
\includegraphics[width=0.5\textwidth]{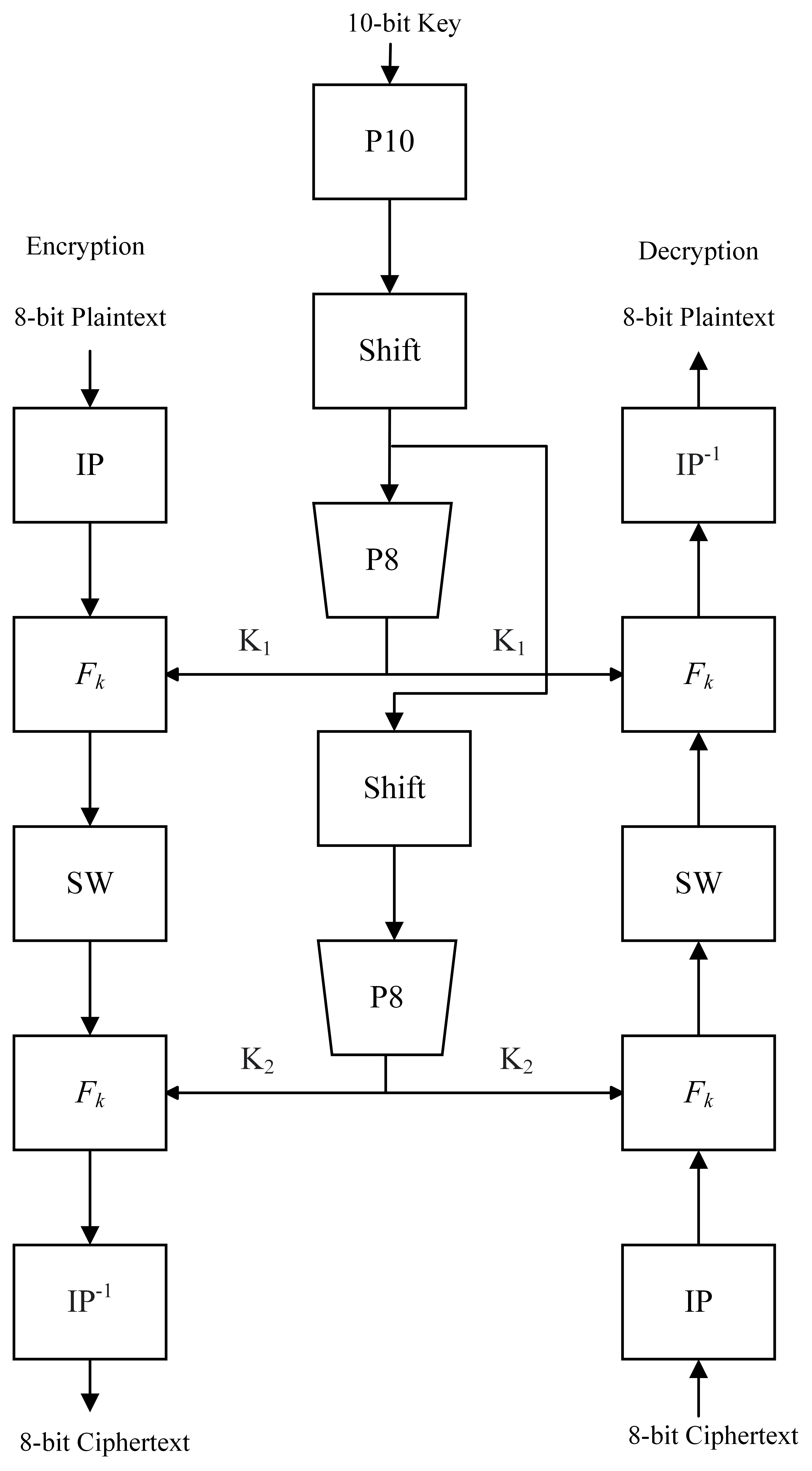}
\caption{Simplified DES (SDES)}
\label{Fig:Simplified DES}
\end{figure}

 In order to prevent an adversary from reaching his goals, some security measures are Applied to cryptosystems, namely confidentiality, data integrity, authentication, and non-repudiation.

\begin{enumerate}
\item Confidentiality means the transmitted message or information is kept secret, and only the authorized parties have the means to decipher the information.
\item Data integrity makes sure that the messages are not being modified. This stops the adversary from reaching their third goal.
\item Authentication helps Bob to correctly identify the sender as Alice, thus stopping the adversary from posing as Alice.
\item Non-repudiation prevents Alice from denying she sent the message.
\end{enumerate}

  Cryptographic algorithms are gathered under two main branches; symmetric algorithms and asymmetric algorithms. In symmetric algorithms both Alice and Bob have the same key. Since the communication channel is insecure, this key must be previously decided on through secure ways. The encryption and decryption keys are either the same, or very similar that the decryption key can easily be derived from the encryption key. But sometimes Alice and Bob cannot agree on a key beforehand. They could be very far away from each other and cannot get together to determine a secret key, and there may not be a secure way for Alice to send Bob the secret key. She cannot just send Bob a secret key through any open channel, because an adversary can interrupt the channel and get their hands on the key. Thus making the key useless. To get around this problem asymmetric algorithms, usually called public key algorithms, are used. In public key algorithms each party has their key pairs, one public and one private key. As can be understood from their names, private keys are kept secret, and public keys can be known by everyone. The public key is computed from the private key in a way that finding the private key from the public key is infeasible. Alice encrypts the message she wants to send using Bob’s public key. The message can only be decrypted with the corresponding private key, which only Bob has. Therefore Alice can send a secret message even though they are far away and cannot decide on a common key together \cite{pub1999data}.\par
Further, the details of the DES cipher is given in the next chpater \cite{han1996improved}.
\section{Data Encryption Standard}
Developed in 1974 by IBM in cooperation with the National Securities Agency (NSA), DES has been the worldwide encryption standard for more than 20 years. For these 20 years, it has held up against cryptanalysis remarkably well and is still secure against all but possibly the most powerful adversaries. Because of its prevalence throughout the encryption market, DES \cite{article} is an excellent interoperability standard between different encryption equipment. The predominant weakness of DES is its 56-bit key which, more than sufficient for the time period which it was developed \cite{nie2009study}, has become insufficient to protect against brute-force attacks modern computers \cite{yazdeen2021fpga}. As a result of the need for a greater encryption strength, DES evolved into triple-DES \cite{coppersmith1996proposed}.

\begin{figure}[h]
\centering
\label{enc & dec}
\includegraphics[width= 0.5\textwidth]{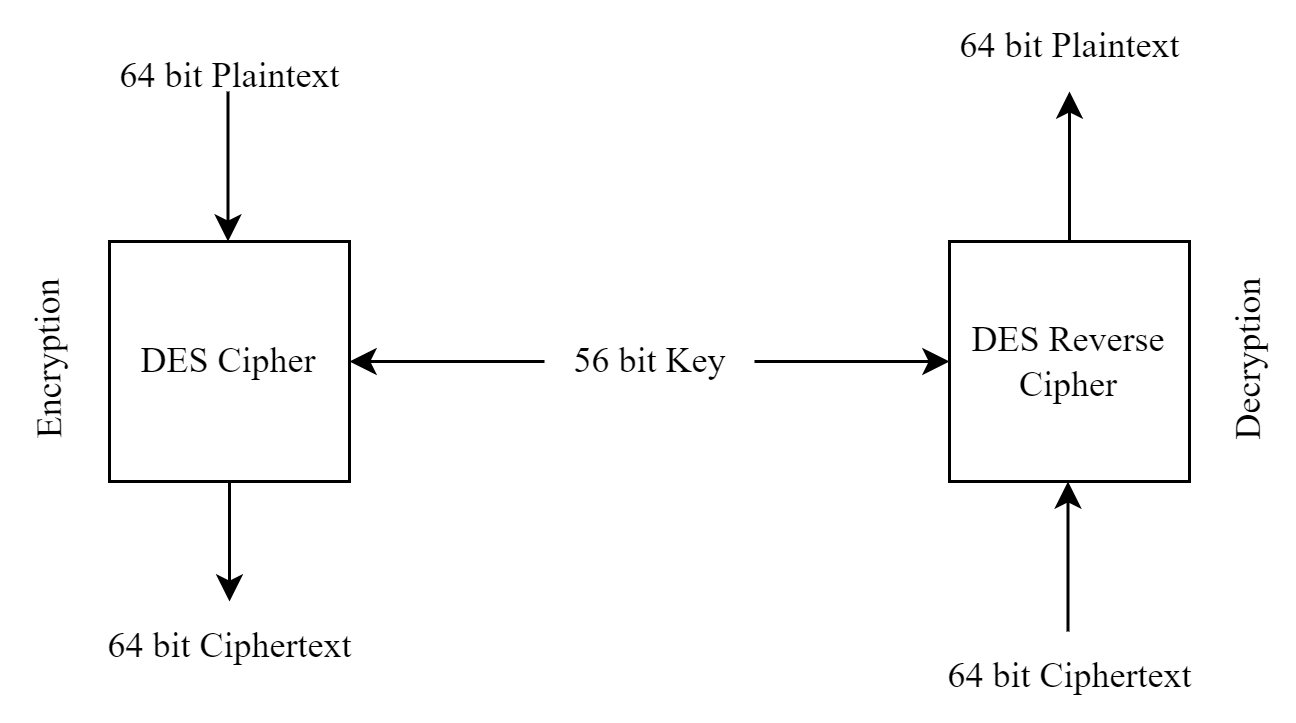}
\caption{Encryption and Decryption}\hspace*{\fill}
\end{figure}

\graphicspath{ {des/} }
\section{DES Encryption}
The Data Encryption Standard is a Feistel cipher. In which round function consists of an expansion, a bitwise XOR-operation XOR operation round key, an S-box layer and a permutation [3]. In encryption n scheme, there are two inputs to the encryption function \cite{thakur2011aes}. the plaintext to be encrypted and the key. In this case, the plaintext must be 64 bits in length and the key is 56 bits in length \cite{coppersmith1994data}.

On the left-hand side of the figure, we can see that the plaintext processing proceeds in three phases. First, the 64-bit plaintext passes through an initial permutation ($IP$) that rearranges the Bits to produce the permuted input \cite{mahajan2013study}. This is followed by a phase consisting of sixteen rounds of the same function, which involves both permutation and substitution functions. The output of the last (sixteenth) round consists of 64 bits that are a function of the input plaintext and the key. The left and right halves of the output are swapped to produce the pre-output. Finally, the pre-output is passed through a permutation [$IP^{-1}$], inverse of the initial permutation function, to produce the make ciphertext. With the exception of initial and final permutations.

\begin{figure}[h]
\centering
\includegraphics[width=0.5\textwidth]{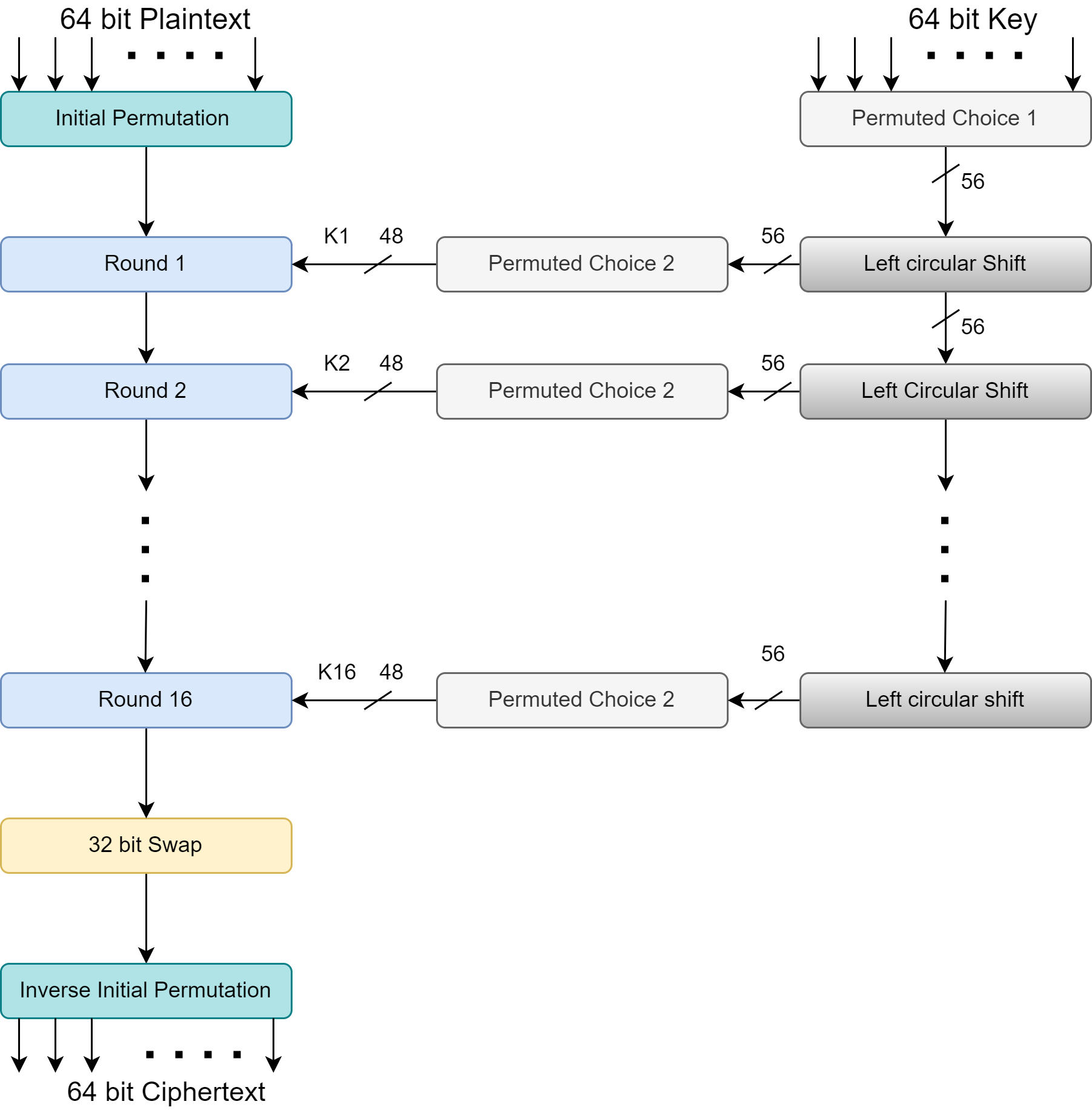}
\caption{Structure of DES}\hspace*{\fill}
\label{structure of DES}
\end{figure}

On the right-hand portion of figure 6-5,6-a-bitkey is used. Initially, the key is passed through a permutation function. Then, for each of the sixteen rounds, a subkey ($K_i$) is produced by the combination f.t Initially, the key is passed through a permutation function. Then, for each of the sixteen rounds, a subkey ($K_i$) is produced by the combination of a left.
\subsection{Initial Permutation and Final Permutation} 
Each of these permutations takes a 64-bit input and permutes them according to a predefined rule. These permutations are keyless straight permutations that are the inverse of each other. For example, in the initial permutation [$IP$], the 58th bit in the input becomes the first bit in the output.
Similarly, in the final permutation [$IP^{-1}$], the first bit in the input becomes the 58th bit in the output. In other words, if the rounds between these two permutations do not exist, the 58th bit entering the initial permutation is the same as the 58th bit leaving the final permutation. The initial permutation is given in \tablename { \ref{Tab:Initial Permutation}}
\begin{figure}[ht]
\centering
\includegraphics[width=0.5\textwidth]{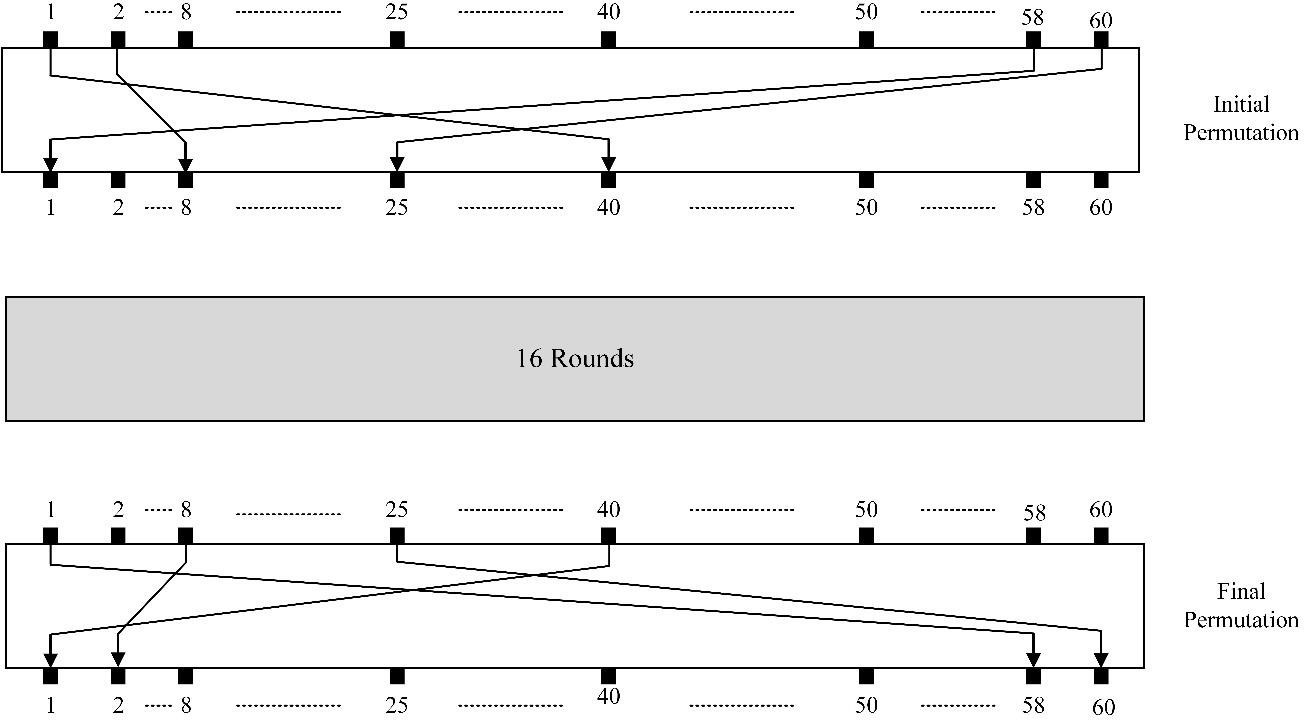}
\caption{Initial and final permutation step in DES}
\label{Initial and final permutation step in DES}
\end{figure}

\begin{table}[ht]
\centering
\caption{Initial Permutation}
\label{Tab:Initial Permutation}
\begin{tabular}{|l|l|l|l|l|l|l|l|}
\hline
58 & 50 & 42 & 34 & 26 & 18 & 10 & 2 \\ \hline
60 & 52 & 44 & 36 & 28 & 20 & 12 & 4 \\ \hline
62 & 54 & 46 & 38 & 30 & 22 & 14 & 6 \\ \hline
64 & 56 & 48 & 40 & 32 & 24 & 16 & 8 \\ \hline
57 & 49 & 41 & 33 & 25 & 17 & 9  & 1 \\ \hline
59 & 51 & 43 & 35 & 27 & 19 & 11 & 3 \\ \hline
61 & 53 & 45 & 37 & 29 & 21 & 13 & 5 \\ \hline
63 & 55 & 47 & 39 & 31 & 23 & 15 & 7 \\ \hline
\end{tabular}
\end{table}

The final permutation is given in \tablename { \ref{Tab:Final Permutation}}.
\begin{table}[ht]
\centering
\caption{Final Permutation}
\label{Tab:Final Permutation}
\begin{tabular}{|l|l|l|l|l|l|l|l|}
\hline
40 & 8 & 48 & 16 & 56 & 24 & 64 & 32 \\ \hline
39 & 7 & 47 & 15 & 55 & 23 & 63 & 31 \\ \hline
38 & 6 & 46 & 14 & 54 & 22 & 62 & 30 \\ \hline
37 & 5 & 45 & 13 & 53 & 21 & 61 & 29 \\ \hline
36 & 4 & 44 & 12 & 52 & 20 & 60 & 28 \\ \hline
35 & 3 & 53 & 11 & 51 & 19 & 59 & 27 \\ \hline
34 & 2 & 42 & 10 & 50 & 18 & 58 & 26 \\ \hline
33 & 1 & 41 & 9  & 49 & 17 & 57 & 25 \\ \hline
\end{tabular}
\end{table}
\section{Rounds}
\begin{figure}[h]
\centering
\includegraphics[width=0.4\textwidth]{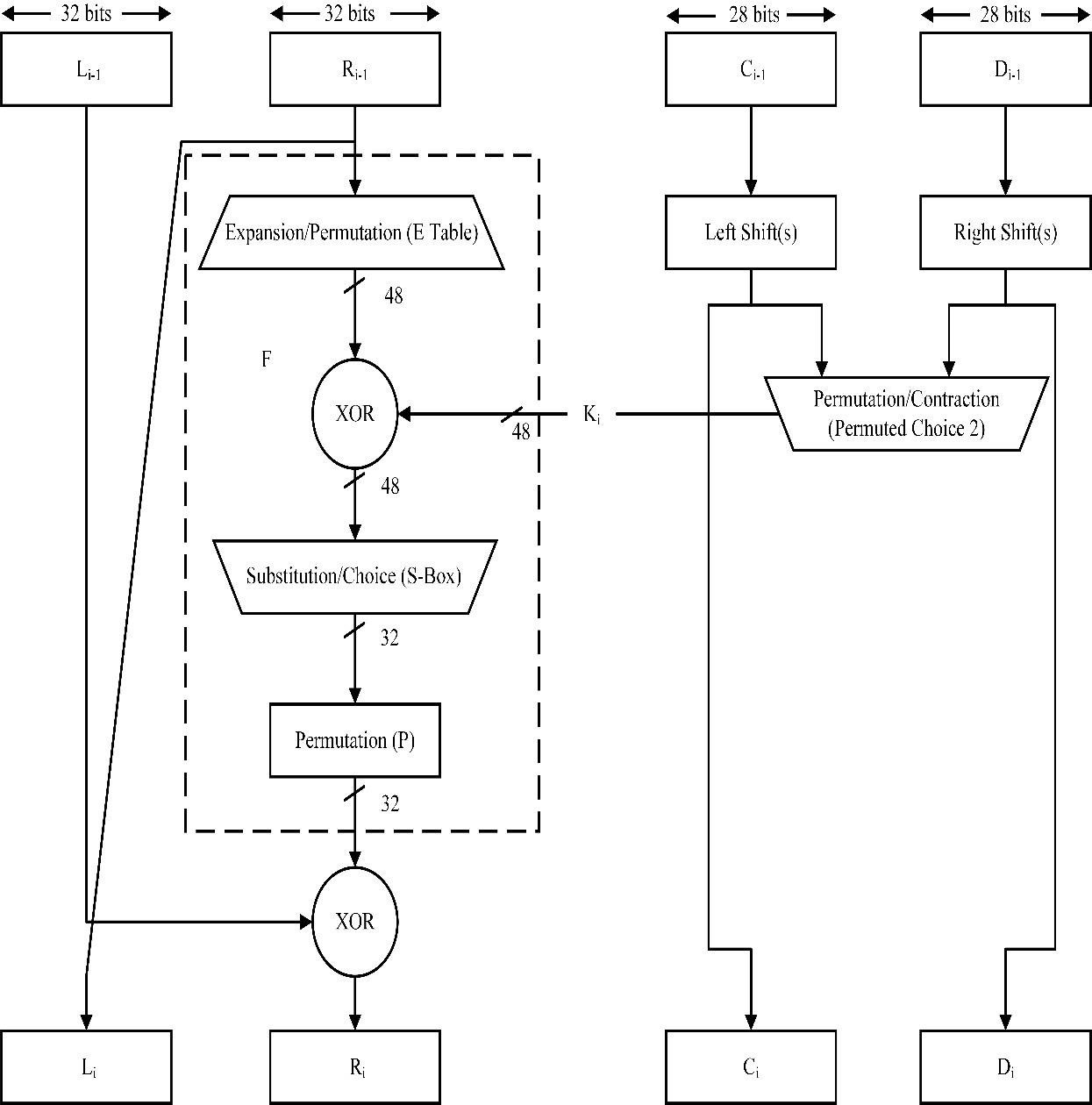} 
\caption{Single round of the DES Algorithm \cite{stallings2006cryptography}.} \hspace*{\fill}
\label{single round of DES Algorithm}
\end{figure}

 DES uses 16 rounds. Each round of DES is a Feistel cipher. \figurename{ \ref{single round of DES Algorithm}} shows the internal structure of a single round. Again, begin by focusing on the left-hand side of the diagram. The left and right halves of each 64-bit intermediate value are treated as separate 32-bit quantities, labeled $L$ (left) and $R$ (right).
As in the Feistel cipher, the overall processing at each round can be summarized in the following formulas:
\\
$L_i$ = $R_i-1$
$R_i$ = $L_i-1$ XOR $FE$($R_i-1$, $K_i$)
\par
The round takes $L_i-1$ and $R_i-1$ from the previous game (or the initial permutation box) and creates $L_i$ and $R_i$, which go to the next round (or final permutation box).

\par
\subsection{Initial Permutation}
A single initial permutation is needed at the beginning of the encryption process. IP is necessary on each block of 64 bits in DES once the entire plaintext has been divided into such blocks. The transposition process goes through this initial permutation. Only once, just before the first round, does the first permutation appear. As seen in the Table I, it provides decisions for how the IP transposition process has to go.
It is possible to claim for example, that the IP replaced the first bit of the original plain-text block with the 58th bit of the original plain-text block, the second bit with the 50th bit of the original plain-text block, etc. This is nothing more than bit shuffling with respect to the original plaintext block.
\subsection{Expansion D-Box}
Since $R_i-1$ is a 32-bit input and KI is a 48-bit key, we first need to expand $R_i-1$ to 48 bits. $R_i-1$ is divided into 8 4-bit sections. Each 4-bit section is then expanded to 6 bits. For each section, input bits 1, 2, 3, and 4 are copied to output bits 2, 3, 4, and 5, respectively. Output bit `1' comes from bit 4 of the previous section; output bit 6 comes from bit 1 of the next section. If sections 1 and 8 can be considered adjacent sections, the same rule applies to bits 1 and 32.

\begin{table}[]
\centering
\caption{Expansion Permutation}
\begin{tabular}{|l|l|l|l|l|l|l|l|}
\hline
32 & 1  & 2  & 3  & 4  & 5  & 64 & 32 \\ \hline
4  & 5  & 6  & 7  & 8  & 9  & 63 & 31 \\ \hline
8  & 9  & 10 & 11 & 12 & 13 & 62 & 30 \\ \hline
12 & 13 & 14 & 15 & 16 & 17 & 61 & 29 \\ \hline
16 & 17 & 18 & 19 & 20 & 21 & 60 & 28 \\ \hline
20 & 21 & 22 & 23 & 24 & 25 & 59 & 27 \\ \hline
24 & 25 & 26 & 27 & 28 & 29 & 58 & 26 \\ \hline
28 & 29 & 30 & 31 & 32 & 1  & 57 & 25 \\ \hline
\end{tabular}
\end{table}

The main part of DES is the DES function. The DES function applies a 48-bit key to the rightmost 32 bits $(R_i-1)$ to produce a 32-bit output. This function is made up of four sections: an expansion D-box, a whitener (that adds key), a group of S-boxes, and a straight D-box, as shown in Fig 6.

\begin{figure}[h]
\centering
\label{DES function}
\includegraphics[height=0.5\textwidth]{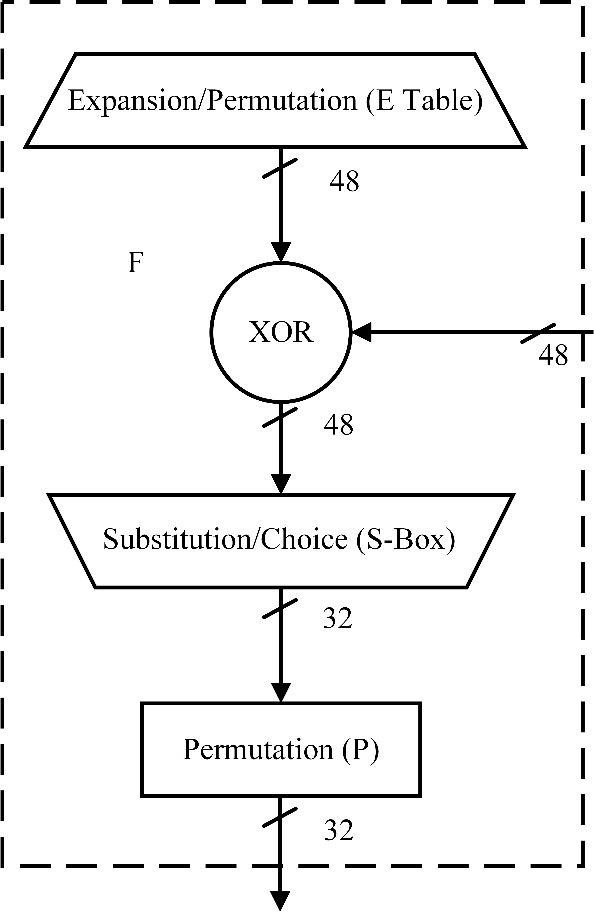}
\caption{DES function}\hspace*{\fill}
\end{figure} 

\subsection{Whitener (XOR)}
After the expansion permutation, DES uses the XOR operation on the expanded right section and the round key. It XORed expansion permutation and key input and gives 48-bit input to s-boxes. Note that both the right section and the key are 48-bits in length \cite{bhatia2016study}.
\subsection{S-Boxes}
The S-boxes do the real mixing (confusion). DES uses 8 S-boxes, each with a 6-bit input and a 4-bit output.
\begin{figure}[h]
\centering
\label{S-box}
\includegraphics[width=0.5\textwidth]{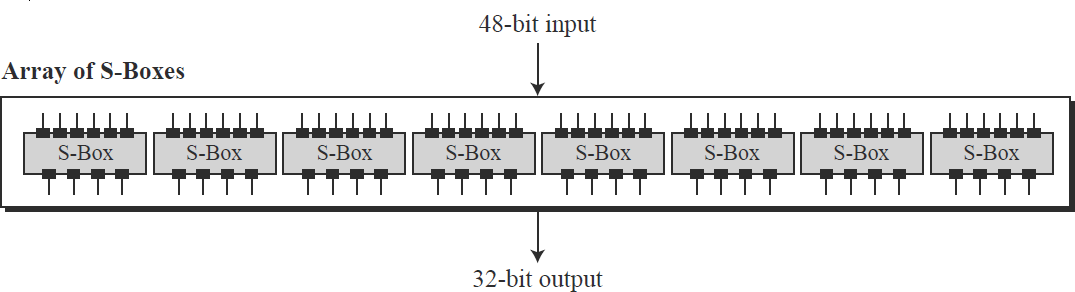}
\caption{S-box}\hspace*{\fill}
\end{figure} 
The 48-bit data from the second operation is divided into eight 6-bit chunks, and each chunk is fed into a box \cite{mohamed2014study}. The result of each box is a 4-bit chunk; when these are combined the result is a 32-bit text. The substitution in each box follows a pre-determined rule based on a 4-row by 16- column table.
\subsection{Final Permutation}
The last operation in the DES function is a permutation with a 32-bit input and a 32-bit output. The input/output relationship for this operation is shown in Table II.
\begin{table}[]
\centering
\caption{Straight Permutation Table}
\begin{tabular}{|l|l|l|l|l|l|l|l|}
\hline
16 & 07 & 20 & 21 & 29 & 12 & 28 & 17 \\ \hline
01 & 15 & 23 & 26 & 05 & 18 & 31 & 10 \\ \hline
02 & 08 & 24 & 14 & 32 & 27 & 03 & 09 \\ \hline
19 & 13 & 30 & 06 & 22 & 11 & 04 & 25 \\ \hline
16 & 17 & 18 & 19 & 20 & 21 & 60 & 28 \\ \hline
20 & 21 & 22 & 23 & 24 & 25 & 59 & 27 \\ \hline
24 & 25 & 26 & 27 & 28 & 29 & 58 & 26 \\ \hline
28 & 29 & 30 & 31 & 32 & 1  & 57 & 25 \\ \hline
\end{tabular}
\end{table}

\section{Examples of DES} 

Let \textit{M} be the plain text message \textit{M} = 0123456789ABCDEF  where \textit{M} is in hexadecimal (base 16) format. Rewriting M in binary format, we get the 64-bit block of text:\\
\textit{M} = 0000 0001 0010 0011 0100 0101 0110 0111 1000 1001 1010 1011 1100 1101 1110 1111,\\
\textit{R} = 1000 1001 1010 1011 1100 1101 1110 1111,\\
\textit{L} = 0000 0001 0010 0011 0100 0101 0110 0111.
\par
The first bit of \textit{M} is `0'. The last bit is `1'. We read from left to right.
DES operates on the 64-bit blocks using key sizes of 56- bits. The keys are actually stored as being 64 bits long, but every 8th bit in the key is not used (i.e., its numbered 8, 16, 24, 32, 40, 48, 56, and 64). However, we will nevertheless number the bits from 1 to 64, going left to right, in the following calculations. But, as you will see, the eight bits just mentioned get eliminated when we create subkeys.
Let \textit{K} be the hexadecimal key \textit{K} = 133457799BBCDFF1. This gives us binary key (setting 1 = 0001, 3 = 0011, etc., and grouping together every eight bits, of which the last one in each group will be unused) \cite{the_des_algorithm_illustrated}.
\textit{K} = 00010011 00110100 01010111 01111001 10011011 10111100 11011111 11110001 .
\\
\section{Key Generation}
The 64-bit key is permuted according to the following table, $PC^{-1}$. Since the first entry in the table is ``57", this means that the $57^{th}$ bit of the original key $K$ becomes the first bit of the permuted key $K_+$. The $49^{th}$ bit of the original key becomes the second bit of the permuted key. The $4^{th}$ bit of the original key is the last bit of the permuted key. Note only 56 bits of the original key appear in the permuted key.

\begin{table}[ht]
\centering
\caption{Permuted Choice-1}
\begin{tabular}{|l|l|l|l|l|l|l|l|}
\hline
57 & 49 & 41 & 33 & 25 & 17 & 9  & 17 \\ \hline
1  & 58 & 50 & 42 & 34 & 26 & 18 & 10 \\ \hline
10 & 2  & 59 & 51 & 43 & 35 & 27 & 09 \\ \hline
19 & 11 & 3  & 60 & 52 & 44 & 36 & 25 \\ \hline
63 & 55 & 47 & 39 & 31 & 23 & 15 & 28 \\ \hline
7  & 62 & 54 & 46 & 38 & 30 & 22 & 27 \\ \hline
14 & 6  & 61 & 53 & 45 & 37 & 29 & 26 \\ \hline
21 & 13 & 5  & 28 & 20 & 12 & 4  & 25 \\ \hline
\end{tabular}
\end{table}
From the original 64-bit key
\\
$K = 00010011 00110100 01010111 01111001 10011011 \\
10111100 11011111 11110001$
we get the 56-bit permutation
\\
$K_+ = 1111000 0110011 0010101 0101111 0101010 \\
1011001 1001111 0001111$

Now, From the permuted key $K_+$, we get
\\
${C_0}$ = 1111000011001100101010101111 \\
$C_1$ = 1110000110011001010101011111 \\
$C_2$ = 1100001100110010101010111111 \\
$C_3$ = 0000110011001010101011111111 \\
$C_4$= 0011001100101010101111111100 \\
$C_5$ = 1100110010101010111111110000 \\
$D_0$ = 0101010101100110011110001111 \\
$D_1$ = 1010101011001100111100011110\\
$D_2$ = 0101010110011001111000111101 \\
$D_3$ = 0101011001100111100011110101 \\
$D_4$ = 0101100110011110001111010101 \\
$D_5$ = 0110011001111000111101010101 \\
$C_6$ = 0011001010101011111111000011 \\
$D_6$ = 1001100111100011110101010101\\
$C_7$ = 1100101010101111111100001100 \\
$D_7$ = 0110011110001111010101010110 \\
$C_8$ = 0010101010111111110000110011 \\
$D_8$ = 1001111000111101010101011001 \\
$C_9$ = 0101010101111111100001100110 \\
$D_9$ = 0011110001111010101010110011 \\
$C_{10}$ = 0101010111111110000110011001 \\
$D_{10}$ = 1111000111101010101011001100 \\
$C_{11}$ = 0101011111111000011001100101 \\
$D_{11}$ = 1100011110101010101100110011 \\
$C_{12}$ = 0101111111100001100110010101 \\
$D_{12}$ = 0001111010101010110011001111 \\
$C_{13}$ = 0111111110000110011001010101 \\
$D_{13}$ = 0111101010101011001100111100 \\
$C_{14}$ = 1111111000011001100101010101 \\
$D_{14}$ = 1110101010101100110011110001  \\
$C_{15}$ = 1111100001100110010101010111 \\
$D_{15}$ = 1010101010110011001111000111  \\ 
$C_{16}$ = 1111000011001100101010101111 \\
$D_{16}$ = 0101010101100110011110001111 \\




 

\begin{table}[ht]
\centering
\caption{Permuted choice-2}
\begin{tabular}{|l|l|l|l|l|l|l|l|}
\hline
14 & 17 & 11 & 24 & 1  & 5  & 9  & 17 \\ \hline
3  & 28 & 15 & 6  & 21 & 10 & 18 & 10 \\ \hline
23 & 19 & 12 & 4  & 26 & 8  & 27 & 09 \\ \hline
16 & 7  & 27 & 20 & 13 & 2  & 36 & 25 \\ \hline
41 & 52 & 31 & 37 & 47 & 55 & 15 & 28 \\ \hline
30 & 40 & 51 & 45 & 33 & 48 & 22 & 27 \\ \hline
44 & 49 & 39 & 56 & 34 & 53 & 29 & 26 \\ \hline
46 & 42 & 50 & 36 & 29 & 32 & 4  & 25 \\ \hline
\end{tabular}
\end{table}

After we apply the permutation PC-2, it becomes
\\
$ K_1$ = 000110 110000 001011 101111 111111 000111 000001 110010 \\
$K_2$ = 011110 011010 111011 011001 110110 111100 100111 100101 \\
$K_3$ = 010101 011111 110010 001010 010000 101100 111110 011001 \\
$K_4$ = 011100 101010 110111 010110 110110 110011 010100 011101 \\
$K_5$ = 011111 001110 110000 000111 111010 110101 001110 101000 \\
$K_6$ = 011000 111010 010100 111110 010100 000111 101100 101111 \\
$K_7$ = 111011 001000 010010 110111 111101 100001 100010 111100 \\
$K_8$ = 111101 111000 101000 111010 110000 010011 101111 111011 \\
$K_9$ = 111000 001101 101111 101011 111011 011110 011110 000001 \\
$K_{10}$ = 101100 011111 001101 000111 101110 100100 011001 001111 \\
$K_{11}$ = 001000 010101 111111 010011 110111 101101 001110 000110 \\
$K_{12}$ = 011101 010111 000111 110101 100101 000110 011111 101001 \\
$K_{13}$ = 100101 111100 010111 010001 111110 101011 101001 000001 \\
$K_{14}$ = 010111 110100 001110 110111 111100 101110 011100 111010 \\
$K_{15}$ = 101111 111001 000110 001101 001111 010011 111100 001010 \\
$K_{16}$ = 110010 110011 110110 001011 000011 100001 011111 110101

\textbf{Encode each 64-bit block of data}
Applying the initial permutation to the block of text \textit{M}, given previously, we get
\\
$M$ = 0000 0001 0010 0011 0100 0101 0110 0111 1000 1001 1010 1011 1100 1101 1110 1111 \\
$IP$ = 1100 1100 0000 0000 1100 1100 1111 1111 1111 0000 1010 1010 1111 0000 1010 1010

Here the $58^{th}$ bit of \textit{M} is \lq{1}\rq, which becomes the first bit of $IP$. The $50^{th}$ bit of \textit{M} is `1', which becomes the second bit of IP. The 7th bit of \textit{M} is `0', which becomes the last bit of $IP$.

Next, divide the permuted block \textit{IP} into a left half $L_0$ of 32 bit, and a right half $R_0$ of 32 bits.

From $IP$, we get $L_0$ and $R_0$
\newline
$L_0$ = 1100 1100 0000 0000 1100 1100 1111 1111 
\\
$R_0$ = 1111 0000 1010 1010 1111 0000 1010 1010

We now proceed through 16 iterations, for 1$\leq$ n $\leq$ 16, using a function f which operates on two blocks--a data block of 32 bits and a key Kn of 48 bits--to produce a block of 32 bits. Let
+ denote XOR addition, (bit-by-bit addition modulo 2). Then for n going from 1 to 16 we calculate (\ref{eq1}) and (\ref{eq2}).
\begin{equation}
    L_n = R_n-1 \label{eq1}
\end{equation}

\begin{equation}
    R_n = L_{n-1} + f(R_{n-1},K_n) \label{eq2}
\end{equation}

This results in a final block, for \textit{n} = 16, of $L_{16}R_{16}$. That is, in each iteration, we take the right 32 bits of the previous result and make them the left 32 bits of the current step. For the right 32 bits in the current step, we XOR the left 32 bits of the previous step with the calculation $f$.

For $n = 1$, we have
\\
$K_1$ = 000110 110000 001011 101111 111111 000111 000001 110010  \\
$L_1$ = R0 = 1111 0000 1010 1010 1111 0000 1010 1010 \\
$R_1 = L_0 + f(R_0,K_1)$

It remains to explain how the function \textit{f} works. To calculate \textit{f}, first expand each block $R_n-1$ from 32 bits to 48 bits. This is done by using a selection table that repeats some of the bits in $R_n-1$. We'll call the use of this selection table the function E. Thus $E(R_n-1)$ has a 32 bit input block, and a 48 bit output block.
After this, We calculate $E(R_0)$ from $R_0$ as follows:
 \\
$R_0$ = 1111 0000 1010 1010 1111 0000 1010 1010\\
$E(R_0)$ = 011110 100001 010101 010101 011110 100001 010101 010101
Next in the \textit{f} calculation, we XOR the output $E(R_n-1)$ with the key $K_n: K_n + E(R_n-1)$.
For $K_1$, $E(R_0)$, we have
\\
$K_1$ = 000110 110000 001011 101111 111111 000111 000001 110010\\
$E(R_0)$ = 011110 100001 010101 010101 011110 100001 010101 010101\\
$K_1+E(R_0)$ = 011000 010001 011110 111010 100001 100110 010100 100111

To this point we have expanded Rn-1 from 32 bits to 48 bits, using the selection table, and XORed the result with the key Kn. We now have 48 bits, or eight groups of six bits. We now do something strange with each group of six bits: we use them as addresses in tables called "S boxes". Each group of six bits will give us an address in a different S box. Located at that address will be a 4-bit number. This 4-bit number will replace the original 6 bits. The net result is that the eight groups of 6 bits are transformed into eight groups of 4 bits (the 4-bit outputs from the S boxes) for 32 bits total.

Write the previous result, which is 48 bits, in the form:
\\
$K_n + E(R_n-1)$ = $B_1B_2B_3B_4B_5B_6B_7B_8$ \\
where each Bi is a group of six bits.
We now calculate it as \\$S_1(B_1)S_2(B_2)S_3(B_3)S_4(B_4)S_5(B_5)S_6(B_6)S_7(B_7)S_8(B_8)$
where $S_i(B_i)$ refers to the output of the $i^{th}$ S box.

To repeat, each of the functions $S_1, S_2, ..., S_8$, takes a 6-bit block as input and yields a 4-bit block as output. For the first round, we obtain as the output of the eight $S$ boxes:\\
$K_1 + E(R_0)$ = 011000 010001 011110 111010 100001 100110 010100 100111\\
\\
$S_1(B_1)S_2(B_2)S_3(B_3)S_4(B_4)S_5(B_5)S_6(B_6)S_7(B_7)S_8(B_8)$ = 0101 1100 1000 0010 1011 0101 1001 0111\\
 
The final stage in the calculation of \textit{f} is to do a permutation P of the S-box output to obtain the final value of f:
\begin{equation}
    f = P(S_1(B_1)S_2(B_2)...S_8(B_8))
\end{equation}

The permutation P is defined in the following table. P yields a 32-bit output from a 32-bit input by permuting the bits of the input block

\begin{table}[ht]
\centering
\caption{Permutation}
\begin{tabular}{|l|l|l|l|l|l|l|l|}
\hline
16 & 7  & 20 & 21 & 1  & 5  & 9  & 17 \\ \hline
29 & 12 & 28 & 17 & 21 & 10 & 18 & 10 \\ \hline
1  & 15 & 23 & 26 & 26 & 8  & 27 & 09 \\ \hline
5  & 18 & 31 & 10 & 13 & 2  & 36 & 25 \\ \hline
2  & 8  & 24 & 14 & 47 & 55 & 15 & 28 \\ \hline
32 & 27 & 3  & 9  & 33 & 48 & 22 & 27 \\ \hline
19 & 23 & 30 & 6  & 34 & 53 & 29 & 26 \\ \hline
22 & 11 & 4  & 25 & 29 & 32 & 4  & 25 \\ \hline
\end{tabular}
\end{table}

From the output of the eight S boxes:
\\
$S_1(B_1)S_2(B_2)S_3(B_3)S_4(B_4)S_5(B_5)S_6(B_6)S_7(B_7)S_8(B_8)$ = 0101 1100 1000 0010 1011 0101 1001 0111\\
we get,
\\
$f$ = 0010 0011 0100 1010 1010 1001 1011 1011 \\
$R_1 = L_0 + f(R_0 , K_1 )$\\

$R_1$ = 1100 1100 0000 0000 1100 1100 1111 1111
+ 0010 0011 0100 1010 1010 1001 1011 1011
= 1110 1111 0100 1010 0110 0101 0100 0100

In the next round, we will have $L_2$ = $R_1$, which is the block we just calculated, and then we must calculate $R_2 = L_1 + f(R_1, K2)$, and so on for 16 rounds. At the end of the sixteenth round we have the blocks $L_16$ and $R_16$. We then reverse the order of the two blocks into the 64-bit block as shown in equation (\ref{eqn:RL}).

\begin{equation}
    R_{16}L_{16} \label{eqn:RL}  
\end{equation}
 
Now, apply a final permutation IP$^{-1}$ and the output of the algorithm has bit 40 of the preoutput block as its first bit, bit 8 as its second bit, and so on, until bit 25 of the preoutput block is the last bit of the output.

If we process all 16 blocks using the method defined previously, we get, on the 16th round,
\\
$L_16$ = 0100 0011 0100 0010 0011 0010 0011 0100\\
$R_16$ = 0000 1010 0100 1100 1101 1001 1001 0101

We reverse the order of these two blocks and apply the final permutation to
\\
$R_16L_16$ = 00001010 01001100 11011001 10010101 01000011 01000010 00110010 00110100\\
$IP^{-1}$ = 10000101 11101000 00010011 01010100 00001111 00001010 10110100 00000101
which in hexadecimal format is 85E813540F0AB405.

This is the encrypted form of \textit{M} = 0123456789ABCDEF: namely, $C$ = 85E813540F0AB405. Decryption is simply the inverse of encryption, following the same steps as above, but reversing the order in which the subkeys are applied.

\section {Symmetric Ciphers}
If we examine the symmetric ciphers in detail, we can see that symmetric ciphers can be divided into two categories; stream ciphers and block ciphers \cite{biryukov2004block}. \par
Stream ciphers use a key-stream, obtained from the original key, and encrypts the plain-text bit by bit. Encryption is usually done by combining the plain-text bits with the corresponding key-stream bits with an XOR operation. In some cases stream ciphers have some advantages over block ciphers, because there is no error propagation. It means that an error made in one bit of cipher-text during transmission only affects the decryption of that bit and doesn’t affect other bits \cite{burke2000architectural}. 

\par
Block ciphers, on the other hand, take the plain-text bits in blocks. Each block is encrypted with the same encryption function and the cipher-text blocks are produced. When the length of the plain-text is not a multiple of the block size some padding is applied to the plain-text \cite{schubert1999efficient}. This padding is usually done by adding a `1' bit followed by necessary amount of `0' bits. Because the encryption function does not change from one block to another, same blocks of plain-text are encrypted to same blocks of cipher-text. When an adversary captures the cipher-text, they can accurately guess some information about the plaintext by using this property. In order to stop any information leakage, some modes of operation are used. \par

\section {Asymmetric Ciphers} 
While modern symmetric ciphers such as AES are very secure, they have some drawbacks in practicality, namely key distribution problem, and the number of keys \cite{lee2020key}. \par
The key distribution problem occurs when Alice and Bob want to determine a secret key. This would be easy if they can come together and decide, but if they have no means to decide on a key in person, they have to decide on the key through a secure channel \cite{8674278}. Since the communication channel is always assumed to be insecure, because it can be easily hacked, this poses a problem. Even if they can somehow solve this problem, they would be facing another problem, the number of keys \cite{senthilkumar2016review}. If there are n users in a network, and all of the users want to communicate with each other secretly, the number of encryption keys needed would be \(\frac{n * (n-1)}{2}\),and each user would have \(n-1\) key pairs they need to know and keep secret. This becomes exponentially infeasible as the number of people increase. The usage of asymmetric ciphers eliminate these problems. Since every user has a pair of keys, and anything encrypted with a specific public key can only be decrypted with the corresponding private key, Alice and Bob doesn’t need to agree on a secret key together beforehand. In addition, nobody would need to store \(n-1\)  key pairs, they only need to store their own private and public keys, and the number of key pairs needed in the network would be reduced to n \cite{stallings2006cryptography}. Cryptographic protocols can be considered as a third main branch of cryptography, and one of the most important primitives they use is called a hash function. Therefore it would be useful to go over the definition of hash functions. \par
In order to understand how public key algorithms work we can imagine a box \cite{chandra2014comparative}. For Alice to send Bob a secret message, first Bob sends Alice a box with an open padlock, for which he has the key. Alice then can put her message in the box, and lock it with the padlock. When Bob receives the box he simply unlocks the padlock and reads the message \cite{stallings2006cryptography}. Of course there are still some security concerns, for example an adversary can intercept the box and replace the padlock with their own lock, or put their own message in the box and act like Alice. To achieve authentication and to prevent these problems, cryptographers have developed some procedures. \par

\subsection{Modes of Operation} 
There are several modes of operation that can be used when encrypting a plaintext with a block cipher. NIST recommends the usage of 5 modes of operation. \cite{stallings2006cryptography}: 
\begin{itemize}
\item Electronic Codebook (ECB)   
\item Cipher Block Chaining (CBC) 
\item Cipher Feedback (CFB)       
\item Output Feedback (OFB)       
\item Counter (CTR)            
\end{itemize}


    In ECB mode, each block is encrypted and decrypted independently from each other. Because the encryption function does not change, identical blocks of plaintext are encypted to identical blocks of ciphertext \cite{celikel2006parallel}. \par
In CBC mode, the ciphertext of one block is XORed with the plaintext of the next block before the encryption \cite{tan2018identification}. For the first plaintext block an Initialization Vector IV is used.
In CFB mode, ciphertext blocks are encrypted with the encryption function instead of the plaintext blocks. Plaintext blocks are XORed with the results of encryption function to obtain the ciphertext blocks. For the first block an IV is used \cite{mister2006attack}. \par
In OFB mode \cite{huang2013building}, IV is repeatedly encrypted with the encryption function and the results are XORed with the plaintext blocks to obtain ciphertext blocks \cite{iwata2014silc}. \par
In CTR mode, a nonce and counter is encrypted and the result is XORed with the plaintext block \cite{lipmaa2000ctr}. The counter is increased each time. \par


All of these modes while having different advantages also have some disadvantages. For example, some of them have parallelizable encryption and decryption but others don’t. The decision of which modes of operation is to be used should be made based on the desired security and performance levels \cite{7946655}.
\par

\section{Results}
Implementation of DES has been performed using VHDL and the results is shown in \tablename{ \ref{Resource_Utilization}} and \tablename{ \ref{Performance_Matrix }}. 


\begin{table}[H]
\setlength{\tabcolsep}{3.5pt}
\centering
\caption{Performance Matrix of DES on Virtex-7 FPGA Device.}
\label{Performance_Matrix }
\begin{tabular}{|c|c|c|c|} 
\hline
\textbf{\begin{tabular}[c]{@{}c@{}}Operating\\ Frequency (MHz)\end{tabular}} & \textbf{\begin{tabular}[c]{@{}c@{}}Datapath Dalay\\ (nS)\end{tabular}} & \textbf{\begin{tabular}[c]{@{}c@{}}Maximum\\ Frequency (MHz)\end{tabular}} & \textbf{\begin{tabular}[c]{@{}c@{}}Dynamic \\ Power (mW)\end{tabular}} \\ \hline
  \multicolumn{1}{|r|}{100}                                                     & \multicolumn{1}{r|}{1.829}                                             & \multicolumn{1}{r|}{246}                                                   & \multicolumn{1}{r|}{8}                                                 \\ \hline
\end{tabular}
\end{table}

\begin{table}[H]
\centering
\caption{Resource Utilization of DES on Virtex-7 FPGA Device.}
\label{Resource_Utilization}
\begin{tabular}{|c|c|c|}
\hline
\textbf{Slices}         & \textbf{LUTs}            & \textbf{Flip-Flops}      \\ \hline
 \multicolumn{1}{|r|}{69} & \multicolumn{1}{r|}{244} & \multicolumn{1}{r|}{139} \\ \hline
\end{tabular}
\end{table}

\section{Conclusion}
Architecture Exploration of Simplified Data Encryption Standard (SDES) and Data Encryption Standard (DES) has been done. Simplified DES (SDES) was designed for educational purposes only, to help learn about modern cryptanalytic techniques. SDES has similar properties and structure as DES but has been simplified to make it much easier to perform encryption and decryption by hand with pencil and paper. Some people feel that learning SDES gives insight into DES and other block ciphers, and insight into various cryptanalytic attacks against them \cite{merkle1981security}.
In DES, 64-bit input is encrypted and decrypted using 56-bit key. At the encryption site, DES takes a 64-bit plaintext and creates a 64-bit ciphertext; at the decryption site, DES takes a 64- bit ciphertext and creates a 64-bit block of plaintext. The same 56-bit cipher key is used for both encryption and decryption.
Implementation of SDES and DES has been performed using Python 3.7 version and VHDL. During this project I have learned thoroughly about various cryptography techniques and ciphers.

\bibliographystyle{IEEEtran} 
\bibliography{refs} 


%





\ifCLASSOPTIONcaptionsoff
  \newpage
\fi

\begin{IEEEbiography}[{\includegraphics[width=1in,height=1.25in]{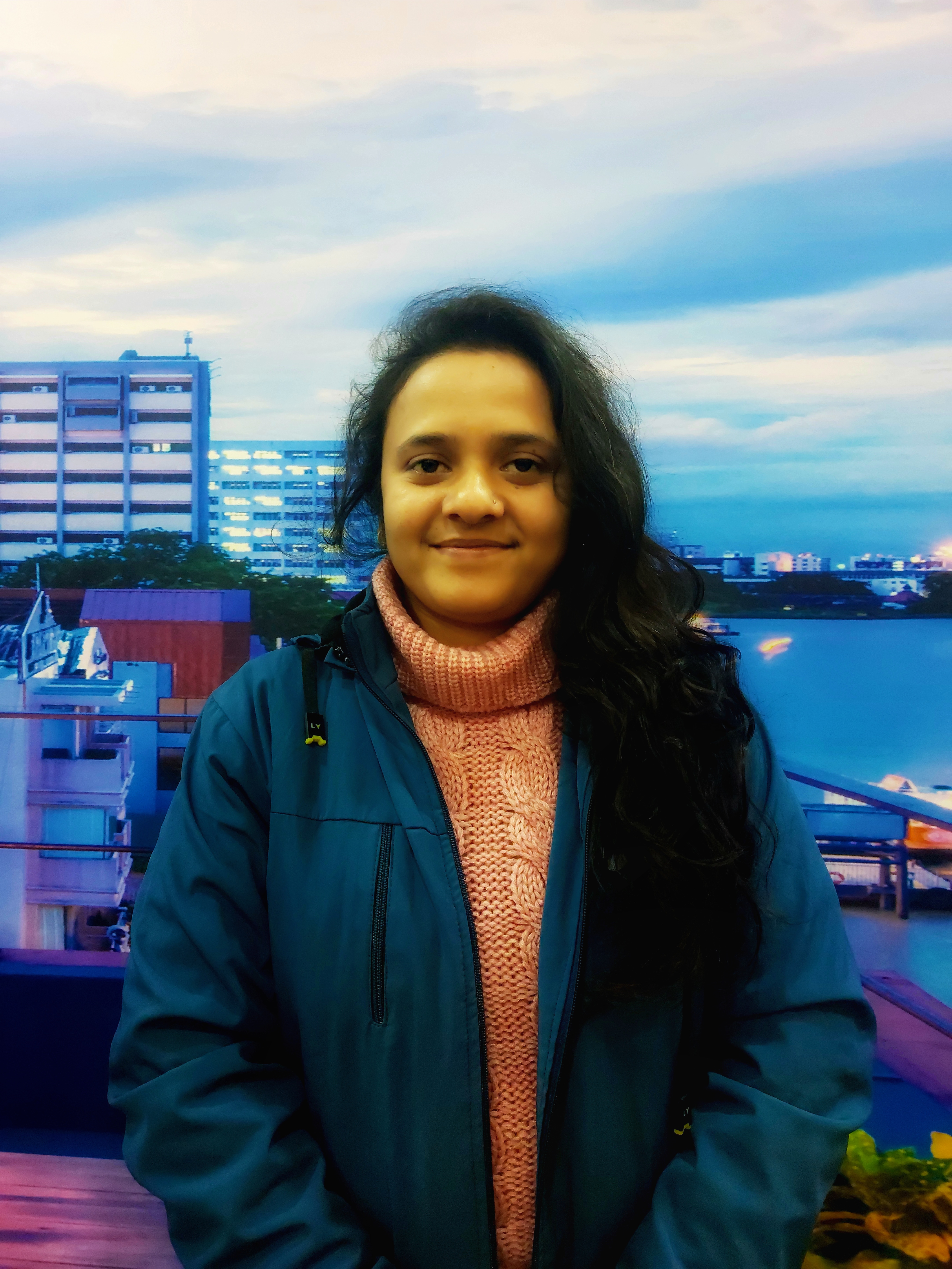}}]
Ruby Kumari, Integrated Dual Degree Ph.D (IDDP) scholar at Academy of Scientific Innovative Research (AcSIR), working area Integrated Circuits and System Group, CSIR-CEERI. Completed B.tech in Electronics and Communication Engineering from Maulana Abul Kalam Azad University of Technology. Her research interest includes Cryptography, Lightweight Ciphers, Digital Logic Design, VLSI Architecture and RTL Design.
\end{IEEEbiography}

\begin{IEEEbiography}[{\includegraphics[width=1in,height=1.25in,clip,keepaspectratio]{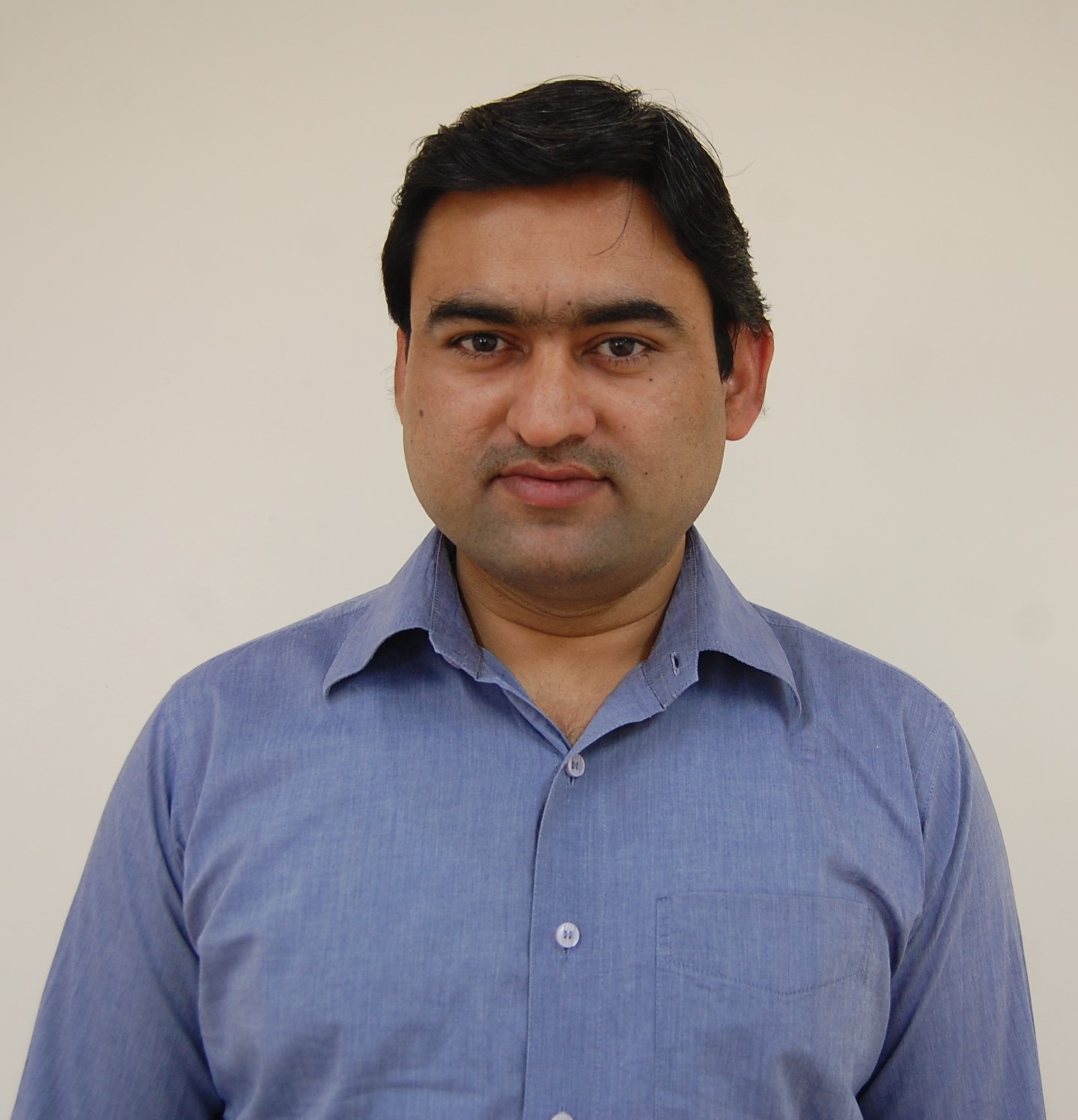}}]{Jai Gopal Pandey} is a Principal Scientist and working in the CSIR-CEERI, Pilani, India since 2005. He is an M.Tech. (Electronics Design and Technology) from U. P. Technical University, Lucknow, in 2003 and a Ph.D. in Electronics Engineering from Birla Institute of Technology and Science (BITS), Pilani, India in 2015. 
\par
His research interests include High-performance Architecture, System-on-chips (SoCs), Embedded Systems, Cryptography, FPGAs, and ASIC designs. Dr. Pandey is a Senior Member of IEEE and an IETE Fellow.
\end{IEEEbiography}

\begin{IEEEbiography}
[{\includegraphics[width=1in,height=1.25in,clip,keepaspectratio]{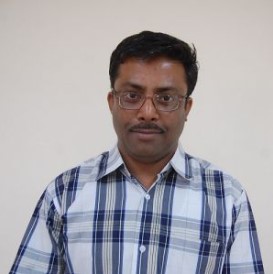}}]{Abhijit Karmakar} received the B.E. degree in Electronics and Telecommunication Engineering in 1993 from Jadavpur University, India, and M.Tech. degree in Electrical Engineering from Indian Institute of Technology (IIT), Madras, India, in 1995. He recieved the Ph.D. degree in  Electrical Engineering from IIT, Delhi, India, in 2007. Since 1995, he has been working with the CSIR - Central Electronics Engineering Research Institute (CEERI), Pilani, India. His research interest span the area of VLSI Design, Signal Processing and related areas.
\end{IEEEbiography}





\end{document}